\documentclass[12pt]{article}
\usepackage{amsfonts,amsmath,amssymb,mathrsfs,url}

\title{The Assumptions of Bell's Proof}

\author{
Roderich Tumulka\footnote{Department of Mathematics, 
     Rutgers University, 110 Frelinghuysen Road, Piscataway, NJ 08854-8019, USA.
     E-mail: tumulka@math.rutgers.edu}
}
\date{January 16, 2015}

\newcommand{\ket}[1]{|#1\rangle}

\newcommand{\va}{\boldsymbol{a}}
\newcommand{\vb}{\boldsymbol{b}}
\newcommand{\vc}{\boldsymbol{c}}
\newcommand{\vd}{\boldsymbol{d}}
\newcommand{\vsigma}{\boldsymbol{\sigma}}
\newcommand{\be}{\begin{equation}}
\newcommand{\ee}{\end{equation}}

\begin{document}
\maketitle
\begin{abstract}
While it is widely agreed that Bell's theorem is an important result in the foundations of quantum physics, there is much disagreement about what exactly Bell's theorem shows. It is agreed that Bell derived a contradiction with experimental facts from some list of assumptions, thus showing that at least one of the assumptions must be wrong; but there is disagreement about what the assumptions were that went into the argument. In this paper, I  make a few points in order to help clarify the situation.

\medskip

\end{abstract}

\section{Introduction}

Different authors have expressed very different views about what Bell's theorem shows about physics. The disagreement concerns particularly the question of which assumptions go into the argument. Since we have to give up one of the assumptions leading to the empirically violated Bell inequality, knowing what the assumptions were is crucial. For example, if author $X$ believes that the argument requires only one assumption, $A_1$, while author $Y$ believes that it requires two, $A_1$ and $A_2$, then $X$ will conclude that $A_1$ must be abandoned, while $Y$ will conclude that either $A_1$ or $A_2$ must abandoned, so $A_1$ may well be true in our world if $A_2$ is false. In this paper, I consider several assumptions that have been mentioned in connection with Bell's theorem, and I look into their roles in the proof of Bell's theorem. While I do not say anything here that has never been said before, I hope that my remarks can nevertheless be helpful to some readers.

For the sake of definiteness, the version of the relevant experiment that I will consider involves two spin-$\frac12$ particles, initially in the singlet spin state $\psi=2^{-1/2}(\left|\uparrow\downarrow\right\rangle - \left|\downarrow\uparrow\right\rangle)$. Two widely separated experimenters, Alice and Bob, have one particle transported to each of them. At spacelike separation, they each choose a direction in space, represented by the unit vectors $\va$ and $\vb$, and carry out a quantum measurement of the spin observable in this direction on their particle, $(\va\cdot \vsigma) \otimes I$ and $I\otimes (\vb\cdot \vsigma)$ with $I$ the $2\times 2$ unit matrix and $\vsigma$ the triple of Pauli matrices. The possible outcomes, $A$ and $B$, are $\pm 1$ on each side. Bell's theorem is the statement that under the assumptions we are about to discuss, Bell's inequality holds; quantum theory predicts violations of Bell's inequality, and experiments have confirmed these predictions.

\section{Local Realism}
\label{sec:realism}

The upshot of Bell's theorem, together with the empirical violation of Bell's inequality, is often described as refuting \emph{local realism}; that is, Bell's theorem is described as requiring, as the two main assumptions, \emph{locality} and \emph{realism}. Let me focus first on realism. Sometimes, different things are meant by this term, so let me formulate several options for what realism might mean here (without implying that ``realism'' is a good name for any of these conditions):

\begin{itemize}
\item[(R1)] Every quantum observable (or at least $(\va\cdot \vsigma) \otimes I$ and $I\otimes (\vb\cdot \vsigma)$ for every $\va$ and $\vb$) actually has a definite value even before any attempt to measure it; the measurement reveals that value.

\item[(R2)] The outcome of every experiment is pre-determined by some (``hidden'') variable $\lambda$.

\item[(R3)] There is some (``hidden'') variable $\lambda$ that influences the outcome in a probabilistic way, as represented by the probability $P(A,B|\va,\vb,\lambda)$.

\item[(R4)] Every experiment has an unambiguous outcome, and records and memories of that outcome agree with what the outcome was at the space-time location of the experiment.
\end{itemize}

I comment on these hypotheses one by one, in the reverse order.

\bigskip

{\bf (R4)} is certainly an assumption required for the proof of Bell's theorem, and was clearly taken for granted by Bell. It would be false in the many-worlds view of quantum theory, as in that view every outcome gets realized.\footnote{One of my readers thought I was implying that the many-worlds view is not a ``realist'' view. That is a misunderstanding, I was not implying that. The conditions (R1) through (R4) are considered here for their potential relevance to Bell's theorem, not for categorizing interpretations of quantum mechanics as realist ones and others.} Bell himself acknowledged that his reasoning does not necessarily apply in a many-worlds framework \cite{Bell86b}:
\begin{quotation}
The `many world interpretation' [\ldots] may have something distinctive to say in connection with the `Einstein Podolsky Rosen puzzle' [i.e., Bell's theorem], and it would be worthwhile, I think, to formulate some precise version of it to see if this is really so.
\end{quotation}
For an analysis of Bell's theorem in the many-worlds framework, I refer the reader to \cite{mw}. In the following, I will leave the many-worlds view aside. Then, it seems that it would be \emph{very} hard to abandon (R4). (R4) is perhaps the mildest form of ``realism'' one could think of: that macroscopic facts are unambiguous, do not change if we observe them, and can be remembered reliably. (R4) would be false if memories and records of outcomes got changed after the fact, regardless of how many people observed the outcome and how the outcomes were recorded. That, of course, would be a veritable conspiracy theory. The upshot is that (R4) is an assumption that many, including myself, are not prepared to abandon. In the following, I will take (R4) for granted.

\bigskip

{\bf (R3)} is really a vacuous assumption, because $\lambda$ in $P(A,B|\va,\vb,\lambda)$ could be just the wave function $\psi$ of the particle pair, a variable which anyhow is well defined in this physical situation and does affect the probability distribution of the outcomes $A,B$. It is \emph{also} allowed that $\lambda$ involves more than just $\psi$, such as particle positions in Bohmian mechanics. But $\lambda=\psi$ is a possibility that is included in the setting of using the expression $P(A,B|\va,\vb,\lambda)$, and thus, (R3) does not commit us to anything like Bohmian particles, that is, to the existence of any variables in addition to $\psi$ (which is what is usually meant by ``hidden variables''). The very notation $P(A,B|\va,\vb,\lambda)$ in the literature on Bell's theorem has presumably suggested to many readers that Bell is making an assumption of hidden variables; after all, $\lambda$ is the hidden variable! But in fact, an assumption of \emph{no hidden variables} is equally allowed by this notation.

\bigskip

{\bf (R2)} is a much stronger assumption than (R3); it is not vacuous, and $\lambda$ could not be taken to be $\psi$ here, as $\psi$ does not determine the outcomes---it only provides probabilities. The assumption (R2) is the reason why Bell's theorem has been called a no-hidden-variables theorem; the idea being that if we have to drop either locality or the assumption of deterministic hidden variables, it would be more attractive to drop the latter and keep locality.

The crucial point about (R2) is that Bell did not make this assumption. The proof of Bell's theorem does not require it. Before I explain why, let me quote what Bell wrote on this point later, in 1981 \cite{Bell81}:
\begin{quotation}
It is important to note that to the limited degree to which {\em determinism} plays a role in the EPR argument [and thus the proof of Bell's theorem], it is not assumed but {\em inferred}. [\ldots]
It is remarkably difficult to get this point across, that determinism is not a {\em presupposition} of the analysis. [emphasis in the original]
\end{quotation}
Now let us look at the role of (R2), i.e., of determinism, in the proof of Bell's theorem. Bell gave two proofs for Bell's theorem, the first in 1964 \cite{Bell64}, the second in 1976 \cite{Bell76}. 
The second one (described also in \cite{Bell81}) is formulated in terms of the expression $P(A,B|\va,\vb,\lambda)$; since it involves a variable $\lambda$ that influences that outcomes $A,B$ in a \emph{probabilistic} way, it becomes clear that it is not assumed that $\lambda$ influences $A,B$ in a \emph{deterministic} way. This illustrates that (R2) is not required for proving Bell's theorem. 

The first proof, of 1964, did not assume (R2) either. And yet, it can easily appear as if (R2) was being assumed, for the following reason. The proof has two parts; part one derives (R2) from the assumption of locality, and part two derives Bell's inequality from 
(R2) and locality. Any reader missing part one will conclude that Bell assumed (R2) \emph{and} locality. And part one may be easy to miss, because it is short, and because all the mathematical work goes into part two, and because part one was, in fact, not developed by Bell but by Einstein, Podolsky, and Rosen (EPR) in their 1935 paper \cite{EPR}. Bell complained later \cite[footnote 10]{Bell81} that people missed part one of the proof:
\begin{quotation}
My own first paper on this subject [i.e., \cite{Bell64}] starts with a summary of the EPR argument \emph{from locality to} deterministic hidden variables. But the commentators have almost universally reported that it begins with deterministic hidden variables.  [emphasis in the original]
\end{quotation}
I will discuss EPR's argument for (R2) from locality in Section~\ref{sec:EPR} below.

\bigskip

{\bf (R1)} implies (R2), and like (R2), it need not be assumed for proving Bell's theorem. In fact, EPR's argument yields (R1) (in the version in parentheses, concerning only the observables $(\va\cdot\vsigma)\otimes I$ and $I\otimes (\vb\cdot \vsigma)$ for all unit vectors $\va,\vb$) if locality is assumed. Thus, the statement (R1) comes up within the 1964 proof of Bell's theorem, but not as an assumption. It does not even come up in Bell's second proof.

\bigskip

If the claims I made are right, one should conclude that there is no \emph{assumption of realism} that enters the proof of Bell's theorem next to the assumption of locality, and thus that we do not have the choice between the two options of abandoning realism and abandoning locality, but that we must abandon locality. One should conclude further that the widespread statement that ``Bell's theorem refutes local realism'' is misleading, and that Bell's theorem simply refutes locality. More detailed discussions of this point can be found in \cite{Nor07,GNTZ,Mau1}.

Bell himself contributed to the misunderstanding that his proof assumed (R2) when he wrote, as the first sentence of the ``Conclusion'' section of his 1964 paper \cite{Bell64}:
\begin{quotation}
In a theory in which parameters are added to quantum mechanics to determine the results of individual measurements, without changing the statistical predictions, there must be a mechanism whereby the setting of one measuring device can influence the reading of another instrument, however remote.
\end{quotation}
This sentence, which would appear like a summary of what is proved in the paper, clearly expresses an \emph{assumption} of determinism as in (R2), and thus may give readers the wrong idea about Bell's theorem. As Norsen \cite{Nor14} has pointed out, Bell may have supposed, when writing this sentence, that his readers were aware that EPR had derived determinism from locality, that therefore it goes without saying that deterministic theories were the only remaining candidates for saving locality, and that therefore a description of the novel contribution of his paper needs to focus only on deterministic theories. Be that as it may, we should not let this sentence mislead us.

\bigskip

Another assumption of Bell's proof is the assumption of effective free will of the experimenters (or, no conspiracy). This assumption, which can be expressed as $P(\lambda|\va,\vb)=P(\lambda)$, means that the experimenter's choices $\va,\vb$ cannot be predicted by the two particles, or conversely, that the particles cannot force the experimenters to make a particular choice of $\va,\vb$. If this were not true, then Bell's inequality could easily be violated by local mechanisms. In this paper, I will take this assumption for granted.

\section{Kolmogorov's Axioms of Probability}

It has been suggested that Bell's proof tacitly assumes the Kolmgorov axioms of probability, and that they can be questioned. Here are a few remarks about this suggestion. The Kolmogorov axioms summarize the usual notion of probability. They say that the events form a Boolean algebra (which means that for any two events $E$ and $F$, also the union $E\cup F$, the intersection $E\cap F$, and the complement $E^c$ are defined and obey the same rules as for sets), that for every event $E$ its probability $P(E)$ is a real number with $0\leq P(E)\leq 1$, and that the $P$ is additive,
\be\label{additive}
P(E\cup F)=P(E) + P(F)
\text{ when }E\cap F=\emptyset\,. 
\ee
(Strictly speaking, $P$ is ``$\sigma$-additive,'' but the difference is not relevant here.)

Clearly, Bell's proof makes use of the Kolmogorov axioms for deriving the Bell inequality. It has been suggested that some of the Kolmogorov axioms might be violated in nature, so that locality may hold true even though the Bell inequality is violated. Specifically, it has been suggested that probabilities can be negative, or that additivity \eqref{additive} can fail. 

However, the $\lambda$ in Bell's proof, which appears in expressions such as $P(A,B|\va,\vb,\lambda)$, is a variable that has a definite value in every run of the experiment. Thus, as we repeat the experiment many times, any value of $\lambda$ occurs with a certain frequency, and it is the limiting value of this frequency that is the relevant notion of probability in Bell's proof. Now, frequencies cannot be negative, and they are necessarily additive, because the number of occurrences of $E\cup F$ equals the number of occurrences of $E$ plus that of $F$, if $E$ and $F$ are disjoint sets in the value space $\Lambda$ of $\lambda$. In fact, since the events are subsets of the set $\Lambda$ of all values of $\lambda$ that occur, the events necessarily form a Boolean algebra. More generally, it lies in the nature of frequencies to obey the Kolmogorov axioms. That is why I see no room for the idea of saving locality by denying the Kolmogorov axioms.

\section{Locality}
\label{sec:loc}

At this point, it seems helpful to look more closely at the locality assumption, which I will call (L). It states:

\begin{itemize}
\item[(L)] If the space-time regions $\mathscr{A}$ and $\mathscr{B}$ are spacelike separated, then events in $\mathscr{A}$ cannot influence events in $\mathscr{B}$.
\end{itemize}

It has been argued about whether ``locality'' is a good name for this statement, and about whether it is an interesting question whether (L) is true. I will, however, leave these debates aside. After all, people can disagree about this and still agree about what Bell's theorem says; and my main concern in this paper is what Bell's theorem says.

In this section, I want to discuss the meaning of the hypothesis (L), at least certain aspects of the meaning. The first question I raise is: In which way, if any, is (L) is different from the statement of \emph{no signaling}? ``No signaling'' means no \emph{superluminal} signaling, of course; that is:

\begin{itemize}
\item[(NS)] If the space-time regions $\mathscr{A}$ and $\mathscr{B}$ are spacelike separated, then it is impossible to transmit a message, freely chosen by an agent (say, Alice) in $\mathscr{A}$, to another agent (say, Bob) in $\mathscr{B}$. 
\end{itemize}

Clearly, (L) implies (NS), as signaling from $\mathscr{A}$ to $\mathscr{B}$ would constitute an instance of an influence from $\mathscr{A}$ to $\mathscr{B}$. And it is a natural thought that (NS) might also imply (L) because if some degree of freedom $x$ in $\mathscr{A}$ could influence some degree of freedom $y$ belonging to $\mathscr{B}$ then Alice might interact with $x$ so as to set it to a desired value encoding part of her message, and Bob might read off that part of the message from $y$. However, this reasoning may not go through if there are \emph{limitations to control} or \emph{limitations to knowledge}, which means that Alice cannot prepare $x$ to have the value she desires, or that Bob cannot find out the value of $y$.

Limitations to control are quite familiar, as we have no control over the random outcome of a quantum experiment, once the initial quantum state of the experiment has been prepared. Limitations to knowledge, in contrast, appear to conflict with the principles of science: It may seem unscientific to believe in the physical existence of a variable that cannot be measured, perhaps like believing in angels. Put differently, if a variable $y$ cannot be measured, it is natural to suspect that $y$ is, in fact, not a well-defined variable---that it does not have a value. However, this conclusion is not necessarily correct, as the following example shows. Suppose that experimenter Carol chooses a direction $\vc$ in 3-space, prepares a spin-$\frac12$ particle with spin up in this direction (i.e., $\vc\cdot\vsigma  \ket{\psi} = \ket{\psi}$ for its quantum state $\ket{\psi}$), and hands it over to Donald, who does not know $\vc$. Donald cannot determine $\vc$ by means of any experiment on the particle, even though there is a fact in nature about what $\ket{\psi}$ and $\vc$ are: After all, Carol, who remembers $\ket{\psi}$ and $\vc$, could tell Donald and predict with certainty that a quantum measurement of $\vc\cdot\vsigma$ will result in the outcome ``up.'' The best Donald can do is to carry out a quantum measurement of $\vd\cdot\vsigma$ in a direction $\vd$ of his choice, which yields one bit of information, and allows a probabilistic statement about whether $\vc$ is more likely to lie in the northern or southern hemisphere with respect to $\vd$, but no more than that. This example shows that there are facts in the world that cannot be revealed by any experiment---a limitation to knowledge about $y=\vc$. 

So, limitations to knowledge are a fact (see \cite{CT13} for further discussion), but nevertheless they are rather unfamiliar, as we normally do not come across them when doing physics. The reason I elaborate on these limitations is that they are relevant to the statement of locality and to Bell's proof. Limitations to knowledge concern the distinction between real and observable; after all, such a limitation occurs when something is real and, at the same time, not observable. This distinction plays a role for reasoning about locality, as (L) refers to \emph{real} events and influences, not to \emph{observable} ones. Let me explain this by means of the ``Einstein boxes example'' (see \cite{Nor05} and references therein).

\section{Einstein's Boxes and Reality}
\label{sec:reality}

Split a 1-particle wave packet into two wave packets of equal size, transport one of them to Paris and the other to Tokyo, and keep each one in a box. At spacelike separation, two experimenters in Paris and Tokyo each apply a detector to their box; quantum mechanics predicts that one and only one of the two detectors clicks, each with probability 1/2. Was any violation of locality involved? Perhaps surprisingly, the answer is: That depends! That is, it depends on what the reality is like. Among the many possibilities for what the reality could be like in this experiment, let me describe two. 

First, the possibility advocated by Einstein, in disagreement with the orthodox interpretation of quantum mechanics, was that the particle actually had a well-defined position before the detection; that is, that the particle actually was already in either Paris or Tokyo, and the detectors merely found out the location and made it known to the experimenters. If that scenario were right, (L) would hold true in this experiment, as the particle traveled slower than light, and the answer of each detector was determined simply by the presence or absence of the particle at the location of the detector, without any influence from the other location. The anti-correlation between the two detection results arises from a common cause in the common past, namely the particle's path towards either Paris or Tokyo. 

Second, the possibility advocated by Bohr was that there is no fact about the particle position before we make a quantum measurement, and the outcome gets generated randomly by nature at the moment of the quantum measurement. If this scenario were right, (L) would be violated in this experiment. After all, in a Lorentz frame in which the Paris experiment occurred first, nature made a random decision about its outcome with 50-50 chances, and this event had to influence, at spacelike separation, the outcome in Tokyo, which always is the opposite of that in Paris. Alternatively, in a Lorentz frame in which the Tokyo experiment occurred first, nature made a random decision about \emph{its} outcome with 50-50 chances, and \emph{this} event had to influence, at spacelike separation, the outcome in Paris.\footnote{\label{fn:nodirection}As a side remark, since relativity suggests that no frame is more ``correct'' than the other, we may fear that a conflict arises here: In a particular run of the experiment, did nature make the random decision in Paris, and did Paris influence Tokyo, or was it the other way around? I find it completely possible that there is no fact of the matter about \emph{where} the random decision was taken, and which direction the influence went. If we give up locality anyway, then the influence does not need to have a direction, and no conflict arises. I will come back to this point at the end of this section.}

This example illustrates that (L) may hold in one scenario and not in another; that is, whether or not (L) holds depends on the scenario. Einstein's and Bohr's scenarios of the boxes example are observationally equivalent, but reality is quite different in the two scenarios, and the truth value of (L) depends on that reality. We may not know which, if any, of the two scenarios is correct, and we may never know. So, we may never know whether (L) is violated in our world when we carry out Einstein's boxes experiment. In order to show that our world is nonlocal, we would have to show that \emph{every} scenario, if it is compatible with the statistics of outcomes predicted by quantum mechanics and confirmed in experiment, has to violate locality. Bell's proof does exactly that.

For deciding whether (L) is valid or violated in a certain scenario, we need to consider the reality according to that scenario. Obviously, each scenario must commit itself to one particular picture of reality; that is, we cannot change our minds in the middle of the reasoning about whether electrons have definite positions or not. Considering reality includes considering reality independently of observation. That is something we often try to avoid in physics, but here we should not avoid it, here exactly that is appropriate and even required. If you think that, in experiments such as the one proposed by Bell, there is no reality independent of observation, then that is one possible scenario, just like the empty set is a set. But you need to consider other scenarios as well. If you are not willing to talk about reality then you cannot talk about locality. If you adopt the positivist attitude that in science one should make only operational statements (of the type ``if we set up an experiment in such-and-such a way, then the possible outcomes are \ldots\ and occur with probabilities \ldots'') then you cannot distinguish between Einstein's and Bohr's scenario, and cannot decide whether locality is valid or violated in them. And then you will not get the point of Bell's theorem.

As a historical remark, it was clear to Einstein that his scenario satisfied locality and Bohr's did not; he used his boxes example for an argument against Bohr's scenario: Einstein believed that locality holds in our world; taking that for granted, it follows that Bohr's scenario is inacceptable, that electrons have definite positions at all times, and that quantum mechanics is incomplete because it leaves these positions out of the description.

One might be worried that the use of (what I called) \emph{scenarios} amounts to adopting classical, rather than quantum, modes of reasoning. Perhaps Werner had in mind something like this when he recently claimed \cite{Wer} that Bell made a tacit assumption by using classical, rather than quantum, probability calculus. Well, the rules of quantum mechanics apply to quantum observables, while classical rules apply to variables that have definite values. One may believe that electrons have definite positions at all times, or that they do not; and likewise with any other observable, one may believe that it has a definite value, or that it does not. In each of these scenarios, the locality condition refers to \emph{events}, i.e., to the unambiguous \emph{facts} about reality in this scenario. If a variable $x$ belonging to space-time region $\mathscr{A}$ has a definite value, that is an event. If $x$ does not have a definite value, then there is no event about the value of $x$. You can, for example, choose Einstein's or Bohr's scenario, in which reality is quite different, and in each scenario it is the reality of that scenario that is relevant to the locality condition. So you need to talk about the facts in this scenario, or, in other words, about those variables that \emph{do} have definite values in this scenario. And to these variables, classical reasoning applies.

\section{Nonlocality and Relativity}

One motivation for doubting that Bell refuted locality arises from the impression that locality is part and parcel of relativity, so that, if we trust in relativity, it is not an option to abandon locality. However, that impression is not right, as it is actually possible to retain relativity and give up locality. This is best demonstrated by specifying a relativistic, nonlocal theory, and one such theory is the \emph{GRW flash theory} \cite{Tum04,Tum06}; see also \cite{Tum07,Mau0} for discussions of this theory, as well as \cite{BDGGTZ} for a different type of relativistic, nonlocal GRW theory. In the GRW flash theory, particles do not have world lines, but they have occasional world points; that is, the reality in this scenario consists of a discrete pattern of material space-time points called ``flashes,'' along with a wave function associated with every spacelike hypersurface. The pattern of flashes is fundamentally random, and the theory prescribes the joint probability distribution of the space-time locations of all flashes as a function of the initial wave function.

This scenario is similar to the one of Bohr described above, in many ways: there are no definite particle positions (at most times); more generally, there are no further (``hidden'') variables in addition to the wave function; the time evolution of a system is stochastic, so randomness is fundamental; the outcome of an experiment gets generated randomly by nature at the moment of the experiment; locality is violated in Einstein's boxes experiment, as well as in Bell's experiment; there is no ``mechanism'' of transporting the influence from $\mathscr{A}$ to $\mathscr{B}$; instead, flashes simply occur randomly with a fundamental probability law that includes Bell's correlation, EPR's, and all others.

Another similar feature is the absence of direction of the nonlocal influence, mentioned in Footnote~\ref{fn:nodirection} above. Let me explain. If locality is violated then there are some situations in which events in $\mathscr{A}$ \emph{must have influenced} events in $\mathscr{B}$, or vice versa. Bell's theorem does not tell us which way the influence went. In the GRW flash theory, there are faster-than-light influences, but there is no fact in nature about which way the influence went---no fact about ``who influenced whom,'' i.e., whether events in $\mathscr{A}$ influenced those in $\mathscr{B}$ or vice versa. In probability theory, two random variables can be correlated, and that is a symmetric relation, without any direction of influence. The situation in the GRW flash theory is exactly the same: the flashes in $\mathscr{A}$ and those in $\mathscr{B}$ are correlated in a nonlocal way, so there is an influence between them without a direction. If you feel that the word ``influence'' necessarily entails a direction, then perhaps this word is not ideal, 
and perhaps ``interaction'' \cite{GNTZ} is a better word. (Of course, ``interaction'' usually means in physics ``a term in the Hamiltonian,'' and here we would have to drop that meaning and return to the original meaning of ``interaction'' in English as mutual effect.)

Bell called locality ``local causality,'' a name with the disadvantage that it adds to the suggestion of direction. After all, a \emph{causal influence} would seem to be one from a cause to an effect, and thus clearly have a direction. That is why I avoid the word ``causal'' in the context of locality.

\section{The EPR Argument}
\label{sec:EPR}

I wish to return to my point of Section~\ref{sec:realism}, that Bell's proof does not assume the hypotheses (R1) or (R2). As I said, this is easily visible in Bell's 1976 proof \cite{Bell76}, while in his 1964 proof \cite{Bell64} it is owed to the EPR argument, which derives (R1) and thus (R2) from locality alone. Let us review this argument. 

The EPR argument, for the experiment involving two spin-$\frac12$ particles in the singlet state, can be put this way: Suppose that Alice and Bob always choose the $z$-direction, $\va=\vb=(0,0,1)$. Quantum mechanics then predicts that the outcomes are perfectly anti-correlated, $A=-B$ with probability 1. Assume locality. Alice's experiment takes place in a space-time region $\mathscr{A}$ and Bob's in $\mathscr{B}$ at spacelike separation. There is a Lorentz frame in which $\mathscr{A}$ is finished before $\mathscr{B}$ begins; thus, in this frame, there is a time at which Alice's experiment already has a definite outcome. She can therefore predict Bob's outcome with certainty, although she cannot transmit this information to Bob before Bob carries out his experiment. Anyway, Bob's outcome was already fixed on some spacelike hypersurface before his experiment. By locality, his outcome was not influenced by events in $\mathscr{A}$, in particular not by whether Alice did any experiment at all. Thus, the state of affairs inside the past light cone of $\mathscr{B}$, but before $\mathscr{B}$ itself, included a fact about the value $B_z$ that Bob will obtain if he carries out a quantum measurement of $I\otimes \sigma_z$. In particular, $B_z$ is a ``hidden variable'' in the sense that it cannot be read off from $\psi$. Since the argument works in the same way for any other direction $\vb$ instead of $z$, there is a well defined value $B_{\vb}$ for every unit vector $\vb$, such that if Bob chooses $\vb_0$ then his outcome will be $B_{\vb_0}$. Since the argument works in the same way for Alice, also her outcome just reveals the pre-determined value $A_{\va}$ for the particular direction $\va$ she chose.

We see how locality enters the argument, and how (R1), and thus (R2), come out. EPR's reasoning is sometimes called a paradox, but the part of the reasoning that I just described is really not a paradox but an argument, showing that (L) implies (R1). The argument is correct, while the premise (L), I argue following Bell, is not. We also see how the argument refers to reality (``events,'' ``state of affairs''), or to different possible scenarios of what the reality might be like, in agreement with my point in Section~\ref{sec:reality} that locality cannot be formulated in purely operational terms.

It may be helpful to make explicit where exactly Bell makes use of the EPR argument in his 1964 paper. On the first page, he writes:
\begin{quotation}
[\ldots] the EPR argument is the following. [\ldots] we make the hypothesis, and it seems one at least worth considering, that if the two measurements are made at places remote from one another the orientation of one magnet does not influence the result obtained with the other. Since we can predict in advance the result of measuring any chosen component of $\vsigma_2$, by previously measuring the same component of $\vsigma_1$, it follows that the result of any such measurement must actually be predetermined. Since the initial quantum mechanical wave function does not determine the result of an individual measurement, this predetermination implies the possibility of a more complete specification of the state.
\end{quotation}
The ``more complete specification of the state'' means the specification of, in addition to $\psi$, the $A_{\va}$ and $B_{\vb}$ for all unit vectors $\va$ and $\vb$, or of some variable $\lambda$ of which all $A_{\va}$ and $B_{\vb}$ are functions.

Actually, the way Bell described the argument in this paper is slightly inaccurate (a point also discussed in \cite[Sec.~III]{Nor14}), as his formulation of the hypothesis expresses not locality but another condition that we may call control locality (CL): 
\begin{itemize}
\item[(CL)] If the space-time regions $\mathscr{A}$ and $\mathscr{B}$ are spacelike separated, then Alice's actions in $\mathscr{A}$ cannot influence events in $\mathscr{B}$.
\end{itemize}
It differs from (L) in that the phrase ``events in $\mathscr{A}$'' has been replaced by ``Alice's actions in $\mathscr{A}$.'' Since Alice's actions in $\mathscr{A}$ are particular events in $\mathscr{A}$, (L) implies (CL). However, while (L) alone implies (R1), (CL) does not, although (CL) and (R2) (i.e., determinism) together imply (R1). The reason why (CL) without (R2) does not imply anything is that, for general stochastic theories, it is not even clear what (CL) should mean. In a deterministic theory, the meaning is clear, as there the outcome $B$ is a function of $\lambda$ and the parameters $\va$ and $\vb$, and (CL) means that this function does not depend on $\va$. In a stochastic theory, however, it is not clear what it should mean that the parameter $\va$ does not influence the random outcome $B$, as it is not clear what the counterfactual statement ``if $\va$ had been different, then $B$ would have been the same'' should mean; if we re-run the randomness, then $B$ may come out differently even for the same $\va$.\footnote{The same difficulty arises with making sense of Conway and Kochen's MIN condition for stochastic theories \cite{GTTZ10}.} So, the passage just quoted from Bell says that (CL) implies (R1), while the correct thing to say is that (L) implies (R1).

To sum up, I have considered various statements that have been claimed to be assumed in the derivation of Bell's inequality and thus to be candidates for being refuted by experimental violations of Bell's inequality. I have given reasons why some assumptions cannot be dropped, why other statements are in fact not assumed in the derivation, and why yet others are indeed refuted.

\bigskip

\noindent\textit{Acknowledgments.}
I wish to thank the many people I had discussions on this subject with. They are people with very different views, and I can only mention a few: J\"urg Fr\"ohlich, Shelly Goldstein, Richard Healey, Federico Holik, George Matsas, Tim Maudlin, Travis Norsen, Daniel Sudarsky, and Reinhard Werner.
The author was supported in part by grant no.\ 37433 from the John Templeton Foundation.

\end{document}